\documentclass[preprint]{aastex}

\input psfig

\def\kpc{{\rm\,kpc}}\def\msun{{\rm\, M}_\odot}
\def\spose#1{\hbox to 0pt{#1\hss}}\def\lta{\mathrel{\spose{\lower 3pt\hbox{$\mathchar"218$}}
     \raise 2.0pt\hbox{$\mathchar"13C$}}}
\def\gta{\mathrel{\spose{\lower 3pt\hbox{$\mathchar"218$}}
     \raise 2.0pt\hbox{$\mathchar"13E$}}}
\def\kms{\,{\rm km\,s}^{-1}}
\def\etal{et al.\ }

\font\fivebmi=cmmib6
\font\sixbmi=cmmib6	\skewchar\sixbmi='177
\font\ninebmi=cmmib10 at 9pt 	\skewchar\ninebmi='177
\newfam\bmifam
\textfont\bmifam=\ninebmi
\scriptfont\bmifam=\sixbmi
\scriptscriptfont\bmifam=\fivebmi

\mathchardef\alpha="710B
\mathchardef\beta="710C
\mathchardef\gamma="710D
\mathchardef\delta="710E
\mathchardef\epsilon="710F
\mathchardef\zeta="7110
\mathchardef\eta="7111
\mathchardef\theta="7112
\mathchardef\iota="7113
\mathchardef\kappa="7114
\mathchardef\lambda="7115
\mathchardef\mu="7116
\mathchardef\nu="7117
\mathchardef\xi="7118
\mathchardef\pi="7119
\mathchardef\rho="711A
\mathchardef\sigma="711B
\mathchardef\tau="711C
\mathchardef\upsilon="711D
\mathchardef\phi="711E
\mathchardef\chi="711F
\mathchardef\psi="7120
\mathchardef\omega="7121
\mathchardef\varepsilon="7122
\mathchardef\vartheta="7123
\mathchardef\varpi="7124
\mathchardef\varrho="7125
\mathchardef\varsigma="7126
\mathchardef\varphi="7127

\def\chaphead{}
\newcount\eqnumber
\def\today{\ifcase\month\or
 January\or February\or March\or April\or May\or June\or
 July\or August\or September\or October\or November\or December\fi
 \space\number\day, \number\year}

\def\eqnam#1{\xdef#1{(\chaphead\the\eqnumber}}

\def\newe{(\hbox{\chaphead\the\eqnumber})\global\advance\eqnumber by 1}
\def\firste{(\hbox{\chaphead\the\eqnumber a})\global\advance\eqnumber by 1}
\def\laste#1{\advance\eqnumber by -1%
	(\hbox{\chaphead\the\eqnumber #1})\advance\eqnumber by 1}

\def\refe#1{\advance\eqnumber by -#1 (\chaphead\the\eqnumber
     \advance\eqnumber by #1 }

\def\i{\relax\ifmmode{\rm i}\else\char16\fi}
\def\e{{\rm e}}
\def\deg{^\circ}             
\def\frac#1#2{{\textstyle{#1\over#2}}}

\def\d{{\rm d}}
\def\dddot#1{\ddot#1\kern-1.4pt\dot{\phantom{#1}}\kern-3pt}

\def\spose#1{\hbox to 0pt{#1\hss}}

\def\=#1{\overline{#1}}

\def\lta{\mathrel{\spose{\lower 3pt\hbox{$\mathchar"218$}}
     \raise 2.0pt\hbox{$\mathchar"13C$}}}
\def\gta{\mathrel{\spose{\lower 3pt\hbox{$\mathchar"218$}}
     \raise 2.0pt\hbox{$\mathchar"13E$}}}

\def\kms{{\rm\,km\,s^{-1}}}
\def\kpc{{\rm\,kpc}}

\def\msun{{\rm\,M_\odot}}

\def\pc{{\rm\,pc}}

\def\annrev #1 #2 {ARA\&A, #1, #2}
\def\aa #1 #2 {A\&A, #1, #2}
\def\aasupp #1 #2 {A\&AS, #1, #2}
\def\aj #1 #2 {AJ, #1, #2}
\def\apj #1 #2 {ApJ, #1, #2}
\def\apjlett #1 #2 {ApJ, #1, #2}
\def\apjsupp #1 #2 {ApJS, #1, #2}
\def\ban #1 #2 {Bull.\ Astron.\ Inst.\ Netherlands, #1, #2}
\def\mn #1 #2 {MNRAS, #1, #2}
\def\nature #1 #2 {Nat, #1, #2}
\def\pasj #1 #2 {PASJ, #1, #2}
\def\pasp #1 #2 {PASP, #1, #2}

\shorttitle{Microlensing and Galactic Structure}
\shortauthors{Binney, Bissantz and Gerhard}

\begin{document}
\title{Is Galactic Structure Compatible with Microlensing Data?}
\author{James Binney}
\affil{Oxford University, Theoretical Physics, Keble Road, Oxford, OX1 3NP,
U.K.}
\author{Nicolai Bissantz and Ortwin Gerhard}
\affil{Astronomical Institute of the University of Basel, Venusstrasse
7, CH-4102 Binningen, Switzerland}


\begin{abstract}
We generalize to elliptical models the argument of Kuijken (1997), which
connects the microlensing optical depth towards the Galactic bulge to the
Galactic rotation curve. When applied to the latest value from the MACHO
collaboration for the optical depth for microlensing of bulge sources, the
argument implies that the Galactic bar cannot plausibly reconcile the
measured values of the optical depth, the rotation curve and the local mass
density. Either there is a problem with the interpretation of the
microlensing data, or our line of sight to the Galactic centre is highly
atypical in that it passes through a massive structure that wraps only a
small distance around the Galactic centre.
\end{abstract}

\section{Introduction}

Searches for gravitational microlensing events, over 500 of which have now
been observed, provide an important constraint on the mass in the inner
Galaxy. Deriving a mass from an observed optical depth is not
straightforward because one usually has only limited knowledge of the
distances to the stars that are lensed, and even less information about the
distances to the deflecting objects.

Kuijken (1997) showed that the {\it minimum\/} mass required to generate a
given optical depth towards the Galactic centre from an axisymmetric
distribution of matter can be determined without any knowledge of the
location of the lenses provided one knows the optical depth to a source on
the Galaxy's symmetry axis. He showed, further, that the
minimum mass required by the then available microlensing data was barely
compatible with the measured rotation curve and local mass density. Here we
generalize Kuijken's argument to the case of elliptical distributions of
matter, which require less matter for a given optical depth.

The apparent magnitude of a red-clump star at the Galactic centre is
relatively bright, so the microlensing optical depth to such objects
can be determined without significant uncertainty due to blending.
Moreover, the red-clump stars must follow the general distribution of
near infra-red light quite closely, because they are part of the
population of evolved stars that are responsible for most of the
Galaxy's near-IR luminosity (McWilliam \& Rich, 1994). Given this, it
proves possible to deduce from the measured mean optical depth of the
clump giants the optical depth to a source on the Galaxy's axis.

From 13 lensing events Alcock et al.\ (1997) inferred $\tau =
3.9^{+1.8}_{-1.2}\times 10^{-6}$ for clump giants centred on $(l,b) =
(2.55\deg, -3.64\deg)$. From a difference image analysis of observations
that include 99 events, Alcock et al.\ (2000b) recently measured
$\tau=2.9^{+0.47}_{-0.45}\times10^{-6}$ for a mix of stars centred on
$(l,b)=(2.68\deg,-3.35\deg)$, and from this measurement deduced for the same
direction $\tau=(3.88\pm0.6)\times10^{-6}$ for bulge sources after making
allowance for the lower optical depth for lensing of disk stars.  Here we
use our results to show that such large values of the optical depth for
microlensing of bulge sources cannot be reconciled with measurements of the
rotation curve and local mass density, even by elliptical Galaxy models. The
paper is organized as follows. In \S2 we rederive Kuijken's result,
generalize it to elliptical systems, and demonstrate that it is applicable
to the measured clump-giant optical depth.  In \S3 we show that axisymmetric
models are very clearly excluded.  In \S4 we show that barred models can be
excluded, albeit with somewhat less confidence.

\section{Lower limits on the Galactic mass}

We determine the minimum mass in stellar objects that is required to
generate a given optical depth $\tau$ towards sources that lie on the
Galaxy's symmetry axis a distance $h\ll R_0$ from the Galactic plane.

The optical
depth to microlensing of a stellar object at distance $s_0$ is
 \begin{equation}\label{basiceq}
\tau={4\pi G\over c^2}\int_0^{s_0}\d s\,\rho_*\hat s,
\end{equation}
where $s$ is the distance to the lens and
\begin{equation}
\hat s=\Big({1\over s_0-s}+{1\over s}\Big)^{-1}.
\end{equation}
 Consider the contribution to $\tau$ from a circular band of mass $M$ and
radius $r$ around the Galactic centre. If we assume that the band's surface
density never increases with distance from the plane, then its mass will be
minimized for a given optical depth when its surface density is constant and
the line-of-sight to the source just cuts its edge. So we take the
band's half-width to be $h(R_0-r)/R_0$, which makes the band's surface
density 
\begin{equation}\label{sigmaMeq}
\Sigma={M\over4\pi rh}\Big(1-{r\over R_0}\Big)^{-1}.
\end{equation}
 Substituting this into equation (\ref{basiceq}) we find the band's optical
depth to be (Kuijken, 1997)
\begin{equation}\label{axieq}
\tau={GM\over c^2 h}
\end{equation}
 independent of radius.  Realistically we must assume that the surface
density of the band falls off with distance from the plane, and if this
decline is exponential with the optimal scale-height ($h[1-r/R_0]$), the
band's mass must be e times that given by equation (\ref{axieq}) for a given optical
depth, while if the vertical density profile is Gaussian with optimal
scale-height ($h[1-r/R_0]$) equation (\ref{axieq}) underestimates the band's
mass by a factor $\sqrt{\pi\e/2}\simeq2.066$. Given the implausibility of
assuming that the band's scale height is optimal, we can safely conclude
that the minimum stellar mass required in a circular band if the observations are
to be met is 
\begin{equation}\label{axieq2}
M_{\rm a}\equiv{2c^2h\over G}\tau=4.2\times\Big({h\over100\pc}\Big)
\Big({\tau\over10^{-6}}\Big)\times10^9\msun.
\end{equation}

Note that the minimum mass estimate (\ref{axieq2}) holds also if the
mass is widely distributed in radius rather than concentrated in a single
band, because we can imagine a radially continuous mass distribution to be
made up of a large number of bands, and we have shown that the optical depth
contributed by each band depends only on its mass when the band is optimally
configured. 

Can one achieve a higher optical depth within a give mass budget by making
the bands elliptical rather than circular? Imagine deforming an initially
circular  band into an
elliptical shape while holding constant the radius $r$ at which the line of
sight to our sources cuts the band. It is straightforward to show that if
the column density through the band to the sources is to be independent of
the band's eccentricity $e$, its mass $M(e)$ must satisfy
 \begin{equation}
M(e)=M(0){1-e^2\cos^2\phi\over\sqrt{1-e^2}},
\end{equation}
 where $\phi$ is the angle between the band's major axis and the Sun--centre
line. For $\phi<\pi/4$, $M(e)$ is a minimum with respect to $e$ at
 \begin{equation}\label{emineq}
e_{\rm min}=\sqrt{2-\sec^2\phi}.
\end{equation}
 For $\phi=20\deg$, a value favoured by Binney, Gerhard \& Spergel
(1997), the optimal
axis ratio is $q_{\rm min}=0.36$ and $M(e)/M(0)=0.64$; for $\phi=15\deg$, we
find $q_{\rm min}=0.27$ and $M(e)/M(0)=0.50$. Hence,
making the bands elliptical realistically reduces the mass required to
generate a given optical depth by at most $50\%$ to
\begin{equation}\label{ellipeq}
M_\e\equiv{c^2h\over G}\tau=2.1\times\Big({h\over100\pc}\Big)
\Big({\tau\over10^{-6}}\Big)\times10^9\msun.
\end{equation}
 Physically, there is an optimum eccentricity because the major axis of the
ellipse grows with $e$, and for large $e$ this growth in scale overwhelms
the reduction in the mass per unit length around the ellipse that is
possible because the line of sight intersects the ellipse ever more
obliquely as $e$ is increased.

\subsection{Applicability to sources not on the axis}

We wish to apply these formulae to sources that are broadly distributed
around the Galactic centre, rather than lying precisely on 
the Galaxy's symmetry axis. How much error will we incur by so doing?

\begin{figure}
\centerline{\psfig{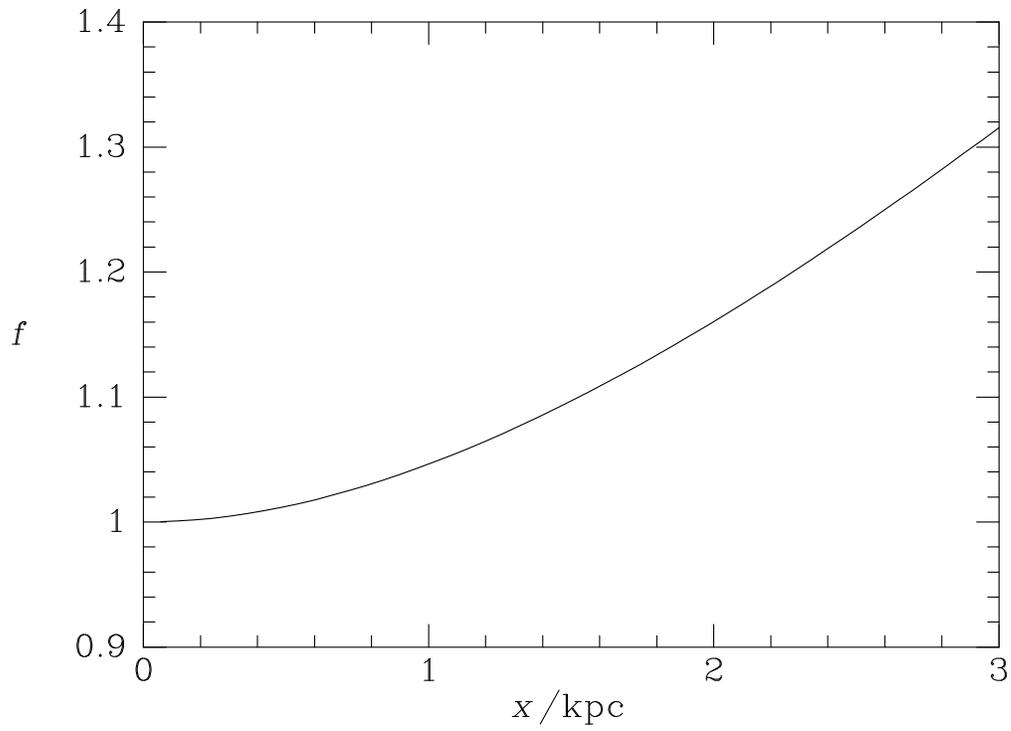}}
\caption{The ratio (\ref{defsfeq}) of optical depths. The abscissa $x$ is the
distance of the sources from the axis. The lenses are distributed such that
their mass density projected along $z$ is proportional to $\e^{-R/R_\d}$
with $R_\d=2.5\kpc$.\label{spread2}}
\end{figure}

Suppose there is a thin circular band of lenses of radius $r$ and imagine
moving two sources along the line of sight away from the axis.  One
source moves toward and one away from the observer, with $x$ being the
distance of each source from the axis. One can easily calculate as a
function of $x$ the average, $\tau(x)$, of the optical depths for each
source to be microlensed by the ring. We then  evaluate the ratio
 \begin{equation}\label{taueq}
{\tau(x,r)\over\tau(0,r)}=\cases{
{\displaystyle {R_0(R_0r-x^2)\over r(R_0^2-x^2)}}&for $x<r$\cr
{\displaystyle {R_0(R_0x-r^2)\over r(R_0-r)(R_0+x)}}&for $x>r$
}
\end{equation}
  For $x<r$, when both sources lie inside the ring, the mean optical depth
is insensitive to $x$ because to first order in $x$ the lower optical depth
to the nearer source is compensated by the higher optical depth to the
further one. For $x>r$, when the sources are outside the ring, the mean
optical depth rises steadily with $x$ because there is nothing to offset
the gain in optical depth for the further source.
To estimate the error involved in assuming that
all sources lie on the axis, we calculate the ratio
 \begin{equation}\label{defsfeq} 
f(x)={\int_{r=0}^{r=R_0}\d\Sigma(r)\tau(x,r)\over
\int_{r=0}^{r=R_0}\d\Sigma(r)\tau(0,r)}
\end{equation}
 of the total mean optical depth due to a series of rings when the sources
lie distance $x$ either side of the axis, and when they lie on axis. In this
calculation we assume that the surface density of each band, $\d\Sigma$, is
given by equation (\ref{sigmaMeq}) with the mass of each band proportional
to $r\d r\,e^{-r/R_\d}$, where $R_\d=2.5\kpc$. Fig.~\ref{spread2} shows the
ratio $f$ as a function of $x$. 

Both the apparent-magnitude distribution of clump giants reported by Stanek
et al.\ (1994) and the COBE/DIRBE near-infrared photometry of the Galaxy
imply that $\sim95\%$ of the clump giants seen towards a typical MACHO field lie
within $2\kpc$ of the point at which the line of sight passes closest to the
Galactic centre (Bissantz \& Gerhard, 2000).\footnote{Moreover, the
apparent-magnitude distribution of the clump giants that have suffered
lensing is consistent with these sources lying within the bulge (Alcock et
al., 1997).}  From
Fig.~\ref{spread2} it now follows that the maximum error made in $\tau$ by
placing any of these sources on the axis is $\sim16\%$. For the $50\%$ of
sources that lie within $\sim700\pc$ of the point of closest approach the
maximum error is $\sim2\%$.

Whereas the non-zero distribution in depth of the clump giants causes us to
slightly underestimate the mean optical depth to them when we place them on
the Galactic axis, the fact that many of the MACHO fields lie at $l\neq0$
gives rise to an error of the opposite sign. Maps of microlensing optical
depth from models based on the COBE/DIRBE data (e.g., Fig.~7 of Bissantz et
al., 1997) show that the errors from this source are also very small.

\section{Lensing by an axisymmetric Galaxy}

We have seen that the mean microlensing optical depth towards bulge sources
at $(l,b)\simeq(2.68\deg,-3.35\deg)$ gives a reasonable approximation to the
optical depth $\tau$ in the formulae above with $h=470\pc$. Setting $\tau$
equal to the  value $(3.88\pm0.6)\times10^{-6}$ inferred by Alcock et al.\
(2000b) we find
 \begin{equation}\label{micros}
M_{\rm a}=(7.6\pm1.2)\times10^{10}\msun\qquad
M_{\rm e}=(3.8\pm0.6)\times10^{10}\msun.
\end{equation}

A crude estimate of the mass required to generate the circular speed
at the Sun is 
\begin{equation}\label{naive}
M_0=(220\kms)^2\times8\kpc/G\simeq8.9\times10^{10}\msun,
\end{equation}
 which is barely more than the minimum stellar mass for an axisymmetric body
from microlensing observations.

The naive estimate (\ref{naive}) is for a spherical mass distribution,
whereas the mass estimates of equation (\ref{micros}) are based on strongly
flattened mass distributions, and $v_c(r)$ for a flattened body can differ
significantly from $v_c(r)$ for the spherical body that has the same
cumulative mass function $M(r)$. Consider therefore Fig.~\ref{vcaxi}, which
shows $v_c(r)$ for two axisymmetric mass distributions in which the vertical
density profile is a Gaussian in $z$ with the scale-height $\sigma$ chosen
to maximize the microlensing optical depth to $(R,z)=(0,470\pc)$ subject to
the auxiliary condition $\sigma\ge30\pc$. The masses of the two bodies are
both $M=7.6\times10^{10}\msun$ to $R=R_0$, so they both produce optical
depth $\tau\sim3.9\times10^{-6}$ to a source at $(R,z)=(0,470\pc)$.  One
body has a surface density $\Sigma(R)\propto\e^{-R/R_\d}$, with
$R_\d=2.5\kpc$, whereas the other has $\Sigma(R)\sim R^{-0.8}$. The rotation
curve of the exponential model clearly exceeds the approximate fit to the
measured circular speed that is given by the short-dashed curve: $v_c(R)\sim
220(R/R_0)^{0.1}\kms$ (Binney et al., 1991). By contrast, the circular-speed
curve of the power-law model lies well below the measured value of $v_c$
even though it corresponds to the same mass profile $M\propto r^{1.2}$. The
reason for this is quite subtle: in this power-law disk there is at any
radius $R$ a substantial outward pull from rings at $R'>R$, and this outward
pull more than compensates for the fact that matter at $R'<R$ exerts a
stronger inward pull than it would if it were distributed spherically. (The
inward and outward pulls cancel exactly for a disk with $\Sigma\sim1/R$.)

Although the power-law disk does not violate the constraint imposed by the
Galaxy's rotation curve, it can be ruled out observationally because it
contributes too much stellar mass to the solar neighbourhood. Specifically, its
volume-density at the Sun is $3.0\msun\pc^{-3}$ compared to the measured
value $0.1\msun\pc^{-3}$ (Cr\'ez\'e et al., 1998; Holmberg \& Flynn, 2000)
and its surface density is $227\msun\pc^{-2}$ compared to an estimated
$\sim35\msun\pc^{-2}$ in stars at $|z|<300\pc$  and the measured
$71\msun\pc^{-2}$ in all matter at
$|z|<1.1\kpc$ (Kuijken \& Gilmore, 1991). Moreover, the power-law model's
large local mass density is vital for its success in evading the
circular-speed limit in that it is intimately connected with the strong
outward pull of material at $R>R_0$.  Hence, if we taper the disk near $R_0$
to avoid conflict with the observed local densities, we will immediately
generate a violation of the limit on the circular speed. A little numerical
experimentation suffices to convince one that there is no way of generating
the required optical depth to $(R,z)=(0,470\pc)$ with a circular disk
without violating either the constraint on the circular speed or that on the
local mass density.

There is, moreover, negligible probability that the Galaxy will be structured
vertically so as to maximize $\tau$ towards $(R,z)=(0,470\pc)$. Hence, we may
be pretty sure that the measured optical depth is not produced by an
axisymmetric Galaxy (Kuijken 1997).

\section{Lensing by a barred Galaxy}

Since it is now generally accepted that the Galaxy contains a bar, the
conclusion we have just reached may not be surprising. Equations
(\ref{micros}) show that non-axisymmetry can in principle reduce the
requirement for stellar mass to half that required in the stellar case. It
is not in practice possible to reduce the required stellar mass by so large
a factor, however, because the structure of the Galaxy's stellar bar is
strongly constrained by both near-IR photometry (Blitz \& Spergel, 1991;
Bissantz et al., 1997) and radio-frequency observations of gas that flows in
the Galactic plane (Englmaier \& Gerhard, 1999; Fux, 1999). We estimate the
reduction in mass that can be achieved as follows.

\begin{figure}
\centerline{\psfig{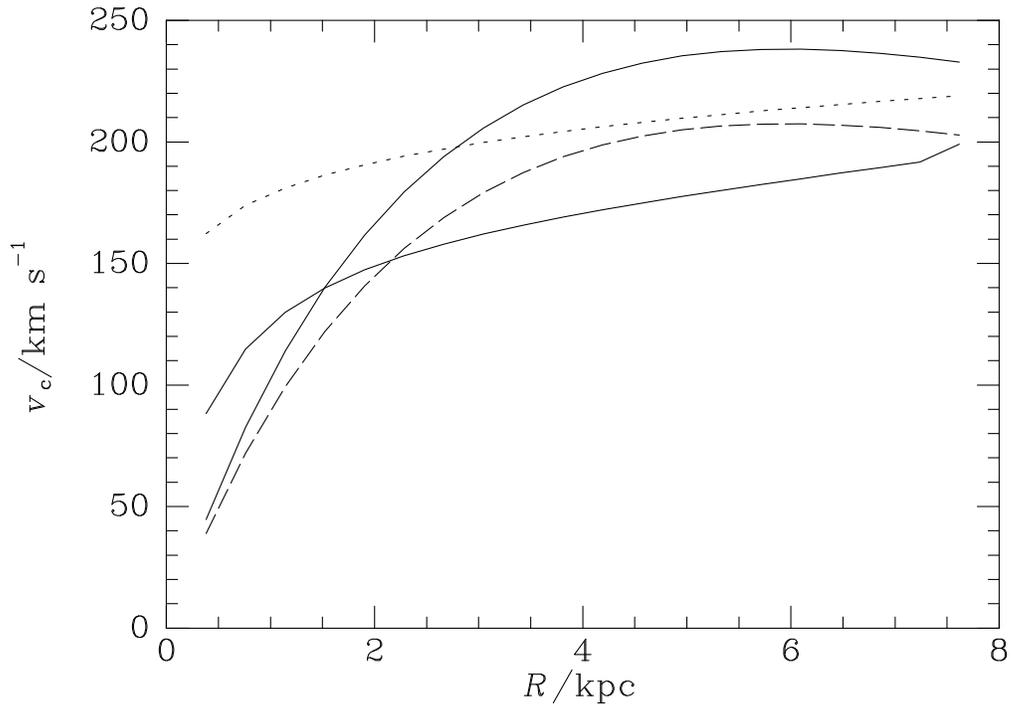}}
\caption{Full curves: the circular speeds of axisymmetric bodies 
that generate $\tau\sim3.9\times10^{-6} $ towards $(R,z)=(0,470\pc)$ with
the minimum mass, $7.6\times10^{10}\msun$ interior to $R_0$.  The curve that
is lower at $R_0$ is for a surface mass density $\Sigma\sim R^{-0.8}$, while
the upper curve is for $\Sigma\sim\e^{-R/R_\d}$ with $R_\d=2.5\kpc$. The
short-dashed curve shows the approximate analytic form $v_c(r)\lta
220(R/R_0)^{0.1}$ (Binney et al., 1991). The long-dashed curve shows the
circular speed of the barred exponential galaxy described in the text.
\label{vcaxi}}
\end{figure}

We suppose that the  surface density of the disk projected along $z$ is
exponential, but that at $R\lta R_0/2$ the material is arranged on
elliptical rings; at every radius the vertical density profile is Gaussian
with the dispersion that maximizes the optical depth for given mass subject
to the condition $\sigma\ge30\pc$.  Then,
the circular speed of the disk will be the same as that of an axisymmetric
disk of the same mass, while the optical depth will be given by
 \begin{equation}\label{Mtoteq}
\tau={GM_{\rm tot}\over2c^2h}\biggl(
{\sqrt{1-e^2}\over1-e^2\cos^2\phi}\,\alpha+(1-\alpha)\biggr),
\end{equation}
where $e$ is determined from equation (\ref{emineq}) and
\begin{equation}
\alpha={1-(1+R_0/2R_\d)\e^{-R_0/2R_\d}\over1-(1+R_0/R_\d)\e^{-R_0/R_\d}}
\end{equation}
 is the fraction of the disk's mass, $M_{\rm tot}$, that lies inside $R_0/2$.
We determine $M_{\rm tot}$ by setting $\tau=3.9\times10^{-6}$ in equation
(\ref{Mtoteq}) and evaluate the resulting curve $v_c(r)$. The long-dashed
curve
in Fig.~\ref{vcaxi} shows the result for $R_\d=2.5\kpc$ and $\phi=20\deg$.
Now the constraint on $v_c$ is not violated, but that on the local
surface density is: the model has $72\msun\pc^{-2}$ at $R_0$.

\begin{figure}
\centerline{\psfig{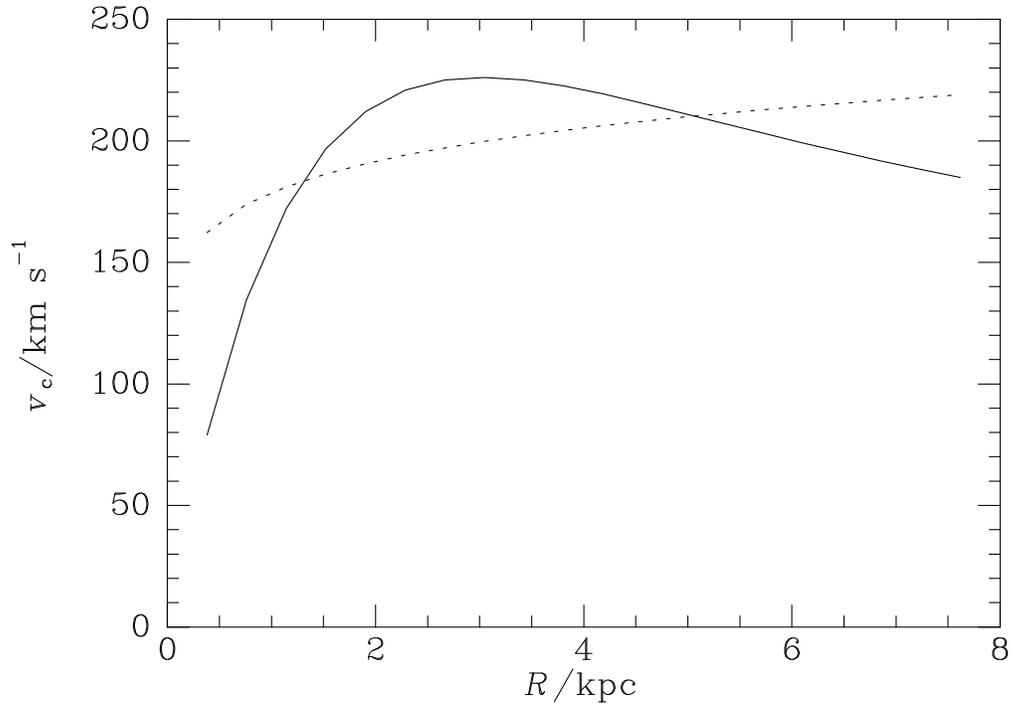}}
\caption{Full curve: the circular speed of the two-exponential elliptical
model described in the text. The dashed curve is the same analytic estimate
of $v_c$ that is shown in Fig.~\ref{vcaxi}.\label{vcbulge}}
\end{figure}

Our assumption that the central Galaxy is barred makes mass placed inside
$R_0/2$ roughly twice as effective at generating optical depth as mass
placed at $R>R_0/2$. Can we then evade both constraints by concentrating
mass in the bar? Fig.~\ref{vcbulge} shows an attempt along these lines. The model
is made up of two exponential components. The larger has a scale length
$R_\d=3\kpc$ and its mass is determined by requiring that at $R_0$ its
surface density is  $35\msun\pc^{-2}$. At $R<R_0/2$ this component has the
eccentricity required by equation (\ref{emineq}) for $\phi=20\deg$. The smaller
component has this same eccentricity throughout, scale-length $R_\d=1\kpc$
and the mass it requires to make the combined optical depth of both
components $3.9\times10^{-6}$. 

In Fig.~\ref{vcbulge} this model's rotation curve has a broad peak around
$v_c\simeq225\kms$ at $R=3\kpc$.  Given the uncertainty in the observed
value of $v_c$ in this region, where the bar is dynamically important, it
might seem reasonable to consider that this barred model is compatible with
both the rotation-curve and the local-density constraints. However,
there are several grounds for caution:

\begin{enumerate}

\item Our models assume that the disk becomes extremely thin near the
Sun and hence violate the Oort limit,
$\rho(R_0)\simeq0.08\msun\pc^{-3}$ (Cr\'ez\'e et al., 1998; Holmberg
\& Flynn, 2000).  We can avoid this problem by imposing $\sigma_{\rm
min}\gta200\pc$ on the scale height, which would significantly reduce
the microlensing efficiency of rings closer to the Sun than
$(\sigma_{\rm min}/h)R_0$. However, the integrated effect can be
small; for example, in the case of the model shown in
Fig.~\ref{vcbulge}, imposing $\sigma_{\rm min}=200\pc$ would reduce
the optical depth by only 3\%.

\item Dynamical models of gas flow in the inner Galaxy imply that the
circular speed rises more steeply at $R\lta500\pc$ than Fig.~\ref{vcbulge}
implies. Specifically, one requires an inner Lindblad resonance at
$R\sim300\pc$ and at $300\gta R\gta30\pc$ dense gas is observed on $x_2$
orbits at speeds that imply $v_c\gta70\kms$. The model for which
Fig.~\ref{vcbulge} is plotted predicts $v_c<70\kms$ at $R<350\pc$ because
it places significant mass at distances $z>400\pc$ from the plane, where it
can generate useful optical depth. To raise $v_c$ at $R<500\pc$ much of this
matter will have to be moved down, closer to the Galactic centre, thus
reducing the model's optical depth. Moreover, near-infrared photometry
confirms that at $R<500\pc$ most of the bulge's mass does lie
below $z=400\pc$.

\item Not all mass causes microlensing. About $5\%$ of the mass of
the inner Galaxy is made up of gas. At $R\gta R_0$ most mass is
contained in a dark halo, most of which does not cause microlensing
(Alcock \etal 2000a). The dark halo is thought to make a significant
contribution to the local circular speed, and its density surely
increases inwards. Hence, the peak circular speed of whatever
component causes microlensing should be strictly less than the
measured value of $v_c$; it is not satisfactory to conclude that it
may be no larger than the measured value.

\item Most importantly, it is not
plausible that the inner Galaxy is structured so as to maximize the optical
depth in a particular direction from a particular star.  In fact mass models
obtained by deprojecting the near-infrared photometry and assuming that mass
follows light are able to reproduce many kinematic properties of the inner
galaxy (Englmaier \& Gerhard, 1999; H\"afner et al., 2000). If mass even
approximately follows light, the photometry implies that the inner
Galaxy is not structured so as to maximize the optical depth produced by a
given mass. In fact the recent model of Bissantz \& Gerhard (2000) yields
$\tau\simeq1.4\times10^{-6}$ in Baade's window.

\end{enumerate}

Thus, we conclude that the marginal satisfaction of the
three principal constraints [optical depth, $v_c$, $\Sigma(R_0)$] that
Fig.~\ref{vcbulge} implies cannot be considered satisfactory and 
that even realistic barred models cannot simultaneously satisfy all the
observational constraints.

\end{document}